\documentstyle[11pt]{article}

\begin{document}
\title{THE TOPOLOGICAL UNITARITY IDENTITIES IN CHERN$-$SIMONS THEORIES}
\author{ V.Ya.FAINBERG\thanks{Permanent address: P.N.Lebedev
Institute of Physics, Moscow}$~$\thanks{Work supported in part by the RFFI
grant 93-02-3379.}$~$ and M.S.SHIKAKHWA \\
Department of Physics, Middle East Technical University\\
06531, Ankara - Turkey\\}
\maketitle
\begin{abstract}
Starting from the generting functional of the theory of relativistic spinors
in $2+1$ dimensions interacting through the pure Chern$-$Simons gauge
field,the S$-$matrix is constructed and seen to be formally the same as that
of spinor quantum electrodynamics in 2+1 dimensions with Feynman diagrams
having external photon lines excluded,
and with the propagator of the topological Chern-Simons photon substituted for
the Maxwell photon propagator.It is shown that the absence of real topological
photons in the complete set of vector states of the total Hilbert space leads
in a given order of perturbation theory to topological unitarity identities
that demand the vanishing of the gauge-invariant sum of the imaginary parts
of
the Feynman diagrams with a given number of internal on-shell free toplogical
photon lines. It is also shown, that these identities can be derived outside
the framework of perturbation theory.The identities are verified explicitly
for the scattering of a fermion-antifermion pair in one-loop order.
\end{abstract}

\newpage
\renewcommand{\thesection}{\Roman{section}}
\section{Introduction}

$~~~~~$In the recent years there has been considerable progress in the
understanding of theories of charged scalar or spinor particles interacting
through the topological Chern-Simons (CS) gauge field .It has been shown that
the addition of the abelian\footnote{In this paper the term CS field will be
understood to mean the abelian CS field unless otherwise stated.}CS kinetic
term to the Lagrangian of quantum electrodynamics (QED) in $2+1$ dimensions,
leads to the appearance of an induced mass of the photon[1].In the work [2],
the important symmetry properties of theories with this topological term have
been investigated,and some one-loop radiative corrections have been
calculatedted.The radiative corrections that the stochastic parameter,i.e
the coefficient multiplying the CS term acquires in the framework of
perturbation
theory have been extensively investigated,and it has been found that for
massive matter,one has only finite corrections at one-loop order,and that
the two-loop $\beta$-function vanishes[3].Later,a theorem was set which states
that there are no corrections to this parameter to all orders of perturbation
theory beyond the finite one-loop one [4].\\
$~~~$On the non-
relativstic domain,it was first proposed by Wilczek that
non-relativistic charged particles interacting through the  topological
CS gauge field can be considered as composite vertices (rigidly bound to a
solenoid which is moving along with them) and were named anyons (particles with
 fractional spin and exotic statistics ).
This idea found wide application,and several
attempts to apply it to the Fractional Quantum Hall Effect,superfluidity
and high temperature superconductivity were made (See the reviews [5] and
references therein for details).\\
$~~~~$In this paper,we are going to consider the theory of relativistic
spinor charged particles coupled to the pure topological CS gauge field.
This theory has the peculiar property that the propagator and many-particle
Green's function of the gauge field are non-zero although
real free particles of the gauge field do not exist
\footnote{The abscence of the
topological photons can be seen most generally from the fact that the CS
term does not contribute to the Hamiltonian because of its independence of
the metric tensor $ g_{\mu \nu}$ in curved space-time.}.\\
$~~~~$We note that if we consider from the beginning the theory of spinor
charged particles coupled to pure CS gauge field,then the appearance of
the parity-conserving  radiative corrections to the CS propagator,being finite,
does not require the introduction of the Maxwell kinetic counter-term for
renormalization.This means that in the framework of perturbation theory,no real
photon appears in this theory.This assertion can be proved by starting
from a Lagrangian that has in addition to the pure CS term ,the Maxwell
term;\begin{equation}
{1\over 4\gamma }\int (F_{\mu \nu }F^{\mu \nu })d^3x
\end{equation}
If one carries out the calculations of any Feynman diagrams of such a
theory and then after renormalization goes to the limit $\gamma \to \infty $,
then one gets the same perturbative results as in the theory with only pure CS
 term (see also part VI ).\\
$~~~~$Our main aim in this paper is to show that the absence of the real
topological photons in the complete set of vector states of the total
Hilbert space of the model,leads to very remarkable topological unitarity
identities.These identities demand that in each order of perturbation theory,
the gauge-invariant sum of the imaginary parts of all Feynman diagrams with
a given number of on-shell internal topological photon lines is equal to zero.
Such identities can also be deduced outside the framework of perturbation
theory. \\
$~~~~$In part II we introduce the generating functional of the theory.In
part III we construct the S-matrix operator.Part IV demonstrates how to deduce
the unitarity identities from the unitarity condition of the theory,and in
part V these identities are verified explicitly for the specific case of the
scattering of a fermion-antifermion pair in one-loop order.Part VI is devoted
to conclusions and discussion.
\section{The Generating Functional}
$~~~~$We start from the $2+1$ dimensional classical action of spinors coupled
to a gauge field whose action is given by the topological pure CS term
\begin{equation}
S=S_m+S_{cs},
\end{equation}
\begin{equation}
S_m=\int d^3x(\bar \psi (i\partial\!\!\!/ +e A\!\!\!/-m)\psi),
\end{equation}
\begin{equation}
S_{cs}={\mu \over 2}\int d^3x(\varepsilon _{\mu \nu \lambda }A^\mu \partial
^\nu A^\lambda ),
\end{equation}
Here $\psi$ and $\psi ^\dagger $,$(\bar \psi =\psi ^\dagger \gamma _0)$ are
two dimensional Grassmann spinor fields,and
the Dirac matrices in $2+1$ dimensions are taken as[6],
\begin{equation}
\gamma _o=\sigma _o~~~~,~~\gamma _i=i\sigma _i~~~~~i=1,2;
\end{equation}
with the $\sigma 's$ being the Pauli spin matrices.These $\gamma $-matrices
satisfy
\begin{equation}
\{ \gamma _\mu ,\gamma _\nu \} _+=2g_{\mu \nu}~~~~;~~
\gamma _\mu \gamma _\nu =g_{\mu \nu }-i\varepsilon _{\mu \nu \lambda }\gamma
^\lambda ,
\end{equation}
the metric $g_{\mu\nu}$  is defined as
\begin{equation}
A_\mu A^\mu =A_\mu g^{\mu \nu }A_\nu ~~~~~,~g_{\mu \nu }=diag.(1,-1,-1)
\end{equation}
$\mu$ and $e$ are two dimensionless coupling constants.A transformation of the
 gauge field of the form:
\begin{equation}
A_\mu \to A_\mu'=\sqrt{\mu }A_\mu ~~~~~~,~~{e\over \sqrt{\mu }}\to g,
\end{equation}
allows us to have only one coupling constant; $g$ in our theory. Then
\begin{equation}
S=\int d^3x(\bar \psi(i\partial\!\!\!/+g A\!\!\!/-m)\psi+{1\over 2}\varepsilon
_{\mu \nu \lambda }A^\mu \partial ^\nu A^\lambda)
\end{equation}
where we have dropped the prime on $A_\mu$. Such a transformation is possible
on the quantum level due to the finite renormalization of the stochastic
parameter $\mu $ [3].\\
$~~~~$In order to carry out the path integral quantization of the model,we
have to handle the constraints.Noting that we have both first and second class
constraints,then generally we have to apply the Batalian-Fradkin-Vilkovisky
 (BFV) method of quantization[7].However,the triviality of the algebra of
the constraints,i.e the fact that the Poisson Brackets of the second class
 constraints are independent of the fields of the theory,and those of the
first class ones are zero,allows us to use the simpler De-Witt-Fadeev
-Popov method[8],which will give the same result as that of BFV method[9]
\footnote{The details of this proof will be published in another paper.}
The generating functional will then assume the following form
\begin{equation}
Z[\eta ,\bar \eta ,J_\mu ]=({\it const.\/})\int DA_\mu (x)D\psi (x)D\bar \psi
(x)\* \ exp\{ iS_m+iS_{cs}+iS_g+i\int d^3x(\bar \eta \psi+\bar \psi \eta +J_
\mu A^\mu )\},
\end{equation}
$J_\mu$ $ (\eta,\bar \eta )$ are external bosonic (fermionic) sources.
The action in the above expression differs from the classical one only by the
appearence of the gauge-fixing action $S_g$;
\begin{equation}
S_g=\int d^3x({-1\over 2\alpha}(\partial _\mu A^\mu )^2),
\end{equation}
where we have adopted a covariant gauge condition.Integrating Eq.(10) over
the gauge field $A_\mu$,we get\footnote{We emphasize that strictly speaking,
it is necessary to introduce in the exponent of (10) a regularization term,for
 instance, the Maxwell term Eq.(1), and then take the limit $\gamma\to\infty$
after integration over $A_\mu$.}
\begin{eqnarray}
Z[\eta ,\bar \eta ,J_\mu ]={\it const.\/}\int D\psi D\bar \psi
\times \exp \{ {-i\over 2}\int d^3xd^3x'(I^\mu (x)D_{\mu \nu }(x-x')I^\nu (x'))
\nonumber \\ +
\int d^3x(\bar\psi (i\partial\!\!\!/ -m)\psi)+\bar \eta \psi +\bar \psi \eta )
\},
\end{eqnarray}
where
\begin{equation}
I^\mu (x)=J^\mu (x)+g\bar \psi (x)\gamma ^\mu \psi (x),
\end{equation}
and $D_{\mu \nu }(x-x')$ is the free CS propagator,
\begin{equation}
D_{\mu \nu }(x-x')=(-\varepsilon _{\mu \nu \lambda }\partial _x^\lambda +
{\alpha\partial_\mu\partial_\nu \over \Box })\int {d^3k\over (2\pi )^3}
{e^{ik(x-y)}\over k^2+i\epsilon}
\end{equation}
Following the general rules,one can determine from the above generating
functional
 all one and many-particle Green's functions of the theory,and use them
to construct perturbatively the scattering amplitude for on-shell processes.\\
$~~~~$Formally,the generating functional (12) is the same as that of QED if
one replaces the CS photon propagator by the Maxwell propagator.On the other
hand,however,inspite of the existence of nonzero Green's function with two and
more external topological photon lines,the corresponding on-shell matrix
elements should be zero
(!) due to the absence of real topological photons.The consequences of this
observation will be investigated in detail in part IV.
\section{The S-matrix Operator}
$~~~~$Although the generating functional Eq.(15) contains all the
information
 of the theory and can be used to derive the scattering amplitudes,it is
actually more convenient to introduce the S-matrix operator in the theory and
use it for this purpose.
The S-matrix operator of scalar charged particles coupled to
CS gauge field has been constructed in the work [10].In our case we have
\begin{equation}
\hat S=T \exp{iS_{int}(\hat \psi ,\hat {\bar \psi },\hat A_\mu )}
\end{equation}
where
\begin{equation}
S_{int}(\hat \psi ,\hat {\bar \psi },\hat A_\mu )=\colon \int d^3x(g\hat A^\mu
\hat{\bar \psi }\gamma _\mu \hat \psi )\colon ,
\end{equation}
": :" means normal ordering, and
$\hat \psi $ and $\hat {\bar \psi }$ are now field operators in the
interation picture:
\begin{equation}
\hat \psi (x)=\int {d^3p\over (2\pi )}\sqrt{{m\over E_{\bf p}}}[b^\dagger ({\bf
}p)u(p)e^{-ipx} + d^\dagger ({\bf p})v(p)e^{ipx}]
\end{equation}
\begin{equation}
\hat {\bar \psi }(x)=\int {d^3p\over (2\pi)}\sqrt{{m\over E_{\bf p}}}
[b^\dagger ({\bf p})\bar u(p)e^{ipx} + d{(\bf p})\bar v(p)e^{-ipx}]
\end{equation}
where $E_{\bf p}=\sqrt{{\bf p^2}+m^2}$,and $b({\bf p})~ (d({\bf p}))$ and
 $b^\dagger ({\bf p})~ (d^\dagger ({\bf p}))$ are the annihilation and creation
operators of particles(antiparticles) respectively,satisfying the usual
anticommutation relations
\begin{equation}
\{b({\bf p}),b^\dagger ({\bf p'})\}_+=\{d({\bf p}),d^\dagger ({\bf p'})\}_+
=\delta ({\bf p-\bf p'}).
\end{equation}
The orthogonal two-component spinors $u({\bf p})$ and $v({\bf p})$,are
respectively the spinors of the positive and negative energy solutions of the
free Dirac equation in $2+1$ dimensions, and have the properties:
\begin{eqnarray}
u(p)\bar u(p)&=&{p\!\!\!/+m\over 2m}\nonumber \\
v(p)\bar v(p)&=&{p\!\!\!/-m\over 2m}
\end{eqnarray}
Obviously,the operatr $ \hat A_\mu (x)$ can not be expanded in terms of the
creation and annihilation operators in a manner similar to $\hat \psi (x)$
and $ \hat {\bar \psi }(x)$.We can bring it into use
only symbolically with the following property : Only the vacuum expectation
value of
the product and the T-product of an even number of the operators $ \hat A_\mu$
is nonvanishing,and reduces to the vacuum expectation value of terms with the
products and
the T-products of two field operators defined as:
\begin{equation}
\langle 0\vert ~T(\hat A_\mu (x)\hat A_\nu (y))~\vert 0\rangle =-iD_{\mu
\nu }(x-y)
\end{equation}
\begin{eqnarray}
\langle 0\vert ~(\hat A_\mu (x)\hat A_\nu (y))~\vert 0\rangle & = & -iD_{\mu
\nu }^\dagger(x-y) \nonumber \\
& =& -i\int {d^3k\over (2\pi )^3}(\varepsilon_{\mu\nu\lambda}ik^
\lambda+{\alpha k_\mu k_\nu\over (k^2+i\epsilon)})\delta (k^2)\theta (k_0)e^{-i
k(x-y)}
\end{eqnarray}
All the matrix elements of the normal product of any number of the the field
operators $\hat A_\mu(x)$ is equal to zero (by definition !).\\
We now make a key observation: The expression (16) of the S-matrix is formally
identical to that of QED.Therefore,we make the following remarkable statement
: All the Feynman diagrams of our theory are identical to those in QED,
if we replace the Maxwell photon propagator by the topological CS
propogator and exclude all diagrams with on-shell external photon lines.
\section{The Toplogical Unitarity Identities}
$~~~~$Here, we are going to use the S-matrix operator introduced in the
previous section to deduce the unitarity identities. However, strictly speaking
, we have to consider the S-matrix operator with the counter-terms introduced
 for the renormalization procedure in each order of perturbation theory(see
for example reference [2] for the counter-terms). To avoid additional
complications, we will assume that all the matrix elements that we consider
have been renormalized.\\
As we have mentioned above,
the absence of the real CS photons means that the complete set of
vector states in the total Hilbert space of the system does not contain these
topological particles.To investigate the consequences of this fact,we
introduce the $\hat T$-matrix,
\begin{equation}
\hat S=1-i\hat T
\end{equation}
where $ \hat S$ is the S-matrix operator (the energy-momentum conserving
 $\delta$-function has been suppressed).The unitarity of the S-operator leads
to the well-known relation:
\begin{equation}
i(\hat T^\dagger -\hat T)=\hat T\hat T^\dagger =2Im\hat T .
\end{equation}
For arbitrary non-diagonal $(\vert i\rangle \neq \vert f\rangle) $
matrix elements on mass-shell,we can write the two equivalent relations:
\begin{equation}
\langle f\vert ~2Im\hat T~\vert i
\rangle =\langle f\vert ~\hat T\hat T
^\dagger ~ \vert i\rangle ,
\end{equation}
\begin{equation}
\langle f\vert ~2Im\hat T~\vert  i\rangle=\sum_n\langle f\vert ~T ~\vert
 n\rangle ~\langle n\vert ~T^\dagger ~\vert i\rangle.
\end{equation}
where in Eq.(26) we have inserted the complete set of physical states
$\{~\vert n\rangle ~\}$ which does not contain the states of the topological
photon, but only those of the charged particles.
{}From Eq.(26) we see that in a given order of perturbation theory,the
Feynman diagrams that contribute to the imaginary part on the l.h.s can not
have internal on-shell topological photon lines because $\{~\vert n\rangle ~\}$
are physical states.
On the other hand,however,investigating Eq.(25) in the framework
of perturbation theory shows that diagrams with internal on-shell photon
lines do appear.This is because the vacuum expectation value of the product
of an even number of
the symbolic operator $\hat A_\mu$ does not vanish due to the non-zero value of
the imaginary part of the photon propagator(see Eq.(22)).Therefore,demanding
the consistency of Eqs.(25) and (26) leads to the important conclusion
 that in a given order of perturbation theory,the gauge-invariant sum of
the imaginary parts of the Feynman diagrams with a given number of on-shell
free topological
photon lines is equal to zero.The vanishing of this sum of the imaginary parts
does not mean the vanishing of the sum of the real parts.As a rule the sum of
such Feynman diagrams
does not vanish,and will give contribution to the process involved. Each
diagram in this sum will be an analytic function of invariant variables.
It is also important to underline the fact that,although the gauge-invariant
sum of the imaginary parts of the
 diagrams with a given number of on-shell internal photon lines vanishes,
the imaginary part of each diagram will not generally  vanish.
It will vanish only if a diagram is gauge-invariant. This will be
 demonstrated when we consider a specific example below.\\$~~~~$The above
arguments give us novel topological unitarity identities which relate the
imaginary parts of Feynman diagrams with a given number of internal on-shell
free photon lines,whose sum is gauge-invariant. That these identities can be
also deduced outside the framework of perturbation theory will be discussed in
part VI.
\section{One-Loop Fermion-Antifermion Scattering}
Now,we illustrate the unitarity identities in the case of scattering of
a fermion-antifermion
pair in one-loop order.This example is of interest also in the
non-relativistic approximation in connection with the perturbative Aharonov-
Bohm scattering amplitude.The gauge-invariant digrams with internal photon
lines that contribute to the process are shown in figure 1 below.
 The analytic expression for the imaginary part of each of these diagrams
is:
\begin{eqnarray}
A_a={2g^4\over (2\pi )^3}\int d^3kd^3k'\left(\delta ^+(k^2)\delta ^+(k'^2)
\delta (p+q-k-k')G_{\mu\lambda}(k)G_{\nu\sigma}(k') \right .  \nonumber \\
\times\frac{\bar v(q)\gamma ^\nu (p\!\!\!/-
 k\!\!\!/+m)\gamma ^\mu u(p)\bar u(p')\gamma ^\lambda (p'\!\!\!/-k\!\!\!/+m)
\gamma
^\sigma v(q')}{((p-k)^2-m^2+i\epsilon)((p'-k)^2-m^2+i\epsilon)((p'-k)^2-m^2
+i\epsilon)} \left .\right),
\end{eqnarray}
\begin{eqnarray}
A_b={2g^4\over (2\pi )^3}\int d^3kd^3k'\left(\delta ^+(k^2)\delta ^+(k'^2)
\delta
 (p+q-k-k')G_{\mu\lambda}(k)G_{\nu\sigma}(k') \right . \nonumber \\
\times  {\bar v(q)\gamma ^\nu (k\!\!\!/-
q\!\!\!/+m)\gamma ^\mu u(p)\bar u(p')\gamma ^\sigma (p'\!\!\!/-k\!\!\!/+m)
\gamma ^\lambda
v(q')\over ((k-q)^2-m^2+i\epsilon )((p'-k)^2-m^2+i\epsilon)}\left .\right).
\end{eqnarray}
where $G_{\mu \nu }(k)=\varepsilon_{\mu \nu \alpha }{k^\alpha \over (k^2+i
\epsilon )}$ ($\alpha=0$ gauge), and $\delta ^+(k^2)=\delta (k^2)\theta (k_0)$.
We for simplicity,restrict ourselves to the case of forward scattering,in which
case the imaginary parts of these diagrams give contribution
to the total scattering cross-section of the two particles.So,we have now in
the center of mass frame for forward scattering (we hereon suppress the
irrelevant overall constant):
\begin{eqnarray}
A_a=\int d^3kd^3k'\left(\delta ^+(k^2)\delta ^+(k'^2)\delta (p+q-k-k')
\right. G_{\mu\lambda}(k)G_{\nu\sigma}(k')\nonumber \\
\times {(\gamma ^\nu (p\!\!\!/-k\!\!\!/ +m)
\gamma ^\mu)_{qp}(\gamma ^\lambda (p\!\!\!/-k\!\!\!/ +m)\gamma ^\sigma )_{pq}
\over 4(p. k)^2}\left .\right )
\end{eqnarray}
\begin{eqnarray}
A_b=\int d^3kd^3k'\left(\delta ^+(k^2)\delta ^+(k'^2)\delta (p+q-k-k')
\right.G_{\mu\lambda }(k)G_{\nu\sigma }(k')\nonumber \\
\times {(\gamma ^\nu (p\!\!\!/-k\!\!\!/+m)\gamma ^\mu )_{qp}
(\gamma ^\sigma (p\!\!\!/-k\!\!\!/'+m)\gamma ^\lambda)_{pq}\over 4(p.k)(p.k')}
\left .\right).
\end{eqnarray}
where $(...)_{qp}=\bar v(q)(...)u(p)$ .
Noting the symmetry of the integrand in k and $k'$, using the identities (20),
 and taking traces over the $\gamma $-matrices as we proceed,we get after
somewhat lengthy calculations :
\begin{equation}
A_a=-\int d^3k\delta ^+(k^2)(1+{p.k\over m^2}+{q.k\over p.k}),
\end{equation}
\begin{equation}
A_b=\int d^3k\delta ^+(k^2)(1+{p.k\over m^2}+{q.k\over p.k}).
\end{equation}
Then  $A_a=-A_b$, or;
\begin{equation}
A_a+A_b=0
\end{equation}
The same result can be obtained for non-forward scattering too\footnote{The
calcualtions for non-forward scattering will be reported in another work.}.This
result verifies the unitarity identities in one-loop oder for the scattering
of a fermion-antifermion pair.
\section{Conclusions and Discussion}
$~~~~$In conclusion,we first note that the fact that the topological unitarity
identities
do not require the vanishing of the imaginary part of each Feynman diagram
with internal on-shell photon lines,can be easily understood for the diagrams
in figure 1.First of all,each of these diagrams is not gauge-invariant by
itself,only their sum is.Moreover, the three diagrams that are the variations
of the diagram in figure 1.a,
are the different boundary values of a single analytical function of the
two invariant variables $s=(p+q)^2$ and $t=(p-p')^2$ for different values of
these variables(i.e in different channels).\\
The topological unitarity identities give additional possibility to check the
self-consistency of the Chern-Simons theories, and to simplify
the calculation of Feynman diagrams in perturbative
 analysis of the theory; in particular the coefficients multiplying these
diagrams.They also provide additional criteria for correctness and gauge-
invariance of the diagrams that resemble the Ward identities.
\\$~~~~$We have carried out some analogous calculations in the
theory of scalar charged particles interacting through pure CS gauge field.In
this case there are four one-loop diagrams with two internal photon lines that
contribute to the scattering amplitude of two opposittely charged particles.
The sum of the imaginary parts of these diagrams (which is
gauge-invariant) vanishes too.
 We also note that the generalization to the non-abelian case should not
be difficult.\\
$~~~$It is necessary to undeline that the S-matrix {\it operator\/} of the
theory of
charged particles interacting through the CS gauge field (see for example Eq.
(15)) contains non-zero normal-ordered products terms of the on-shell
topological photon operators. However, all the {\it matrix elements\/} of the
normal-ordered products of the topological photon operator are "by definition"
 equal to zero. Due to this property, the matrix elements
$\langle f |T^\dagger
T | i\rangle$ in Eq.(25) contain terms with internal topological $D^+_{\mu\nu}
(x)$ functions (See Eq.(30) ). Thus, a comparative analysis of Eqs.(25) and
(26) non-perturbatively shows that the only difference between them comes from
 these terms. So, we come to the conclusion that these two equations will be
 consistent only in the case when each exact matrix element with a given
number of internal topological lines in Eq.(25) vanishes. This statement is a
 direct generalization of the topological unitarity identities in CS theories
to the non-perturbative case.\\
$~~~$Moreover, the same arguments are also applicable in the
non-relativistic case, and analogous identities can be deduced there too.\\
$~~~$Finally, we note that if we introduce into the action in (10) the Maxwell
term (1), then we get (We do not perform the transformation (7) here) formally
the same generating functional Eq.(12) except that the $D_{\mu\nu}$
propagator is replaced by (in momentum space)
\begin{eqnarray}
D_{\mu\nu}(p)=\gamma\left(\frac{g_{\mu\nu}-p_\mu p_\nu /p^2}{p^2-\mu^2\gamma^2
+i\epsilon}+\frac{i\mu\gamma\varepsilon_{\mu\nu\lambda}p^\lambda}
{(p^2+i\epsilon)
(p^2-\mu^2\gamma^2+i\epsilon)}\right)+\alpha\frac{p_\mu p_\nu}{(p^2+i\epsilon)
^2}
\end{eqnarray}
 In the above propagator, the photon acquires the mass $\mu\gamma $ as we
have mentioned in the introduction. The Maxwell term also plays the role of a
regularization term  and the theory becomes superrenormalizable [6]. Since the
 stochastic
parameter $\mu$ and the charge $e$ acquire only finite corrections due to the
superrenormalizability of the theory, then it is possible to consider two
limiting
procedures (in each order of perturbation theory after renormalization). In the
 case $\gamma\to\infty$ the propagator (34) reduces to the pure CS propagator,
 and we get a renormalizable theory. Thus, in this case the topological
unitarity identities hold. In the case $\mu\to 0,~\gamma$ finite, we obtain
pure
 2+1 dimensional QED propagator, and the unitarity identities no longer hold.
 In the intermediate cases, one can show that the total operator of the gauge
 field in the S-matrix consists of two parts: the operator of the real massive
 photon, and that of the massless topological photon that does not give real
  radiation. Thus, the topological unitarity identities still hold in the
sense that the gauge-invariant sum of the imaginary parts of the Feynman
diagrams with at least one on-shell internal topological line vanishes.\\

Acknowledgments.\\
$~~~$We thank Professor N.K.Pak for helpful discussions and
M.Boz who also checked the vanishing of the sum of the imaginary
parts in the scalar case. One of us (V.F) thanks the directorate of ICTP where
 this work was concluded for hospitality.

\end{document}